\documentclass[aps, twocolumn, showpacs, showkeys,nofootinbib,floatfix]{revtex4-1}

\usepackage{amssymb}
\usepackage{amsmath}
\usepackage{graphicx}
\usepackage{hyperref}
\usepackage{longtable}

\begin{document}

\title{A New Unified Dark Fluid Model and Its Cosmic Constraint}

\author{Lixin Xu}
\email{lxxu@dlut.edu.cn}

\affiliation{Institute of Theoretical Physics, Dalian University of Technology, Dalian,
116024, Peoples Republic of China}

\begin{abstract}
In this paper,  we propose a new unified dark fluid (UDF) model with equation of state (EoS) $w(a)=-\alpha/(\beta a^{-n}+1)$, which includes the generalized Chaplygin gas model (gGg) as its special case, where $\alpha$, $\beta$ and $n$ are three positive numbers. It is clear that this model reduces to the gCg model with EoS $w(a)=-B_s/(B_s+(1-B_s)a^{-3(1+\alpha)})$, when $\alpha=1$, $\beta=(1-B_s)/B_s$ and $n=3(1+\alpha)$. By combination the cold dark matter and the cosmological constant, one can coin a EoS of unified dark fluid in the form of $w(a)=-1/(1+(1-\Omega_{\Lambda})a^{-3}/\Omega_{\Lambda})$. With this observations, our proposed EoS provides a possible deviation from $\Lambda$CDM model when the model parameters $\alpha$ and $n$ deviate from $1$ and $3$ respectively. By using the currently available cosmic observations from type Ia supernovae (SN Ia) Union2.1, baryon acoustic oscillation (BAO) and cosmic microwave background radiation (CMB), we test the viability of this model and detect the possible devotion from the $\Lambda$CDM model. The results show that the new UDF model fits the cosmic observation as well as that of the $\Lambda$CDM model and no deviation is found from the $\Lambda$CDM model in $3\sigma$ confidence level. However, our new UDF model can give a non-zero sound speed, as a contrast, which is zero for the $\Lambda$CDM model. We expect the large structure formation information can distinct the new UDF model from the $\Lambda$CDM model.

\end{abstract}



\maketitle

\section{Introduction}

The cosmic observations from type Ia supernovae (SN Ia) indicate that our Universe is undergoing an accelerated expansion \cite{ref:Riess98,ref:Perlmuter99}. Holding the Einstein's gravity theory, the observed energy component $T^{ons}_{\mu\nu}$ and the geometric structure of our Universe in hand, one can deduce the existence of a remained un-obsered energy component
\begin{equation}
T^{dark}_{\mu\nu}=\frac{1}{8\pi G}G_{\mu\nu}-T^{obs}_{\mu\nu},\label{eq:dark}
\end{equation}  
which is dubbed as a dark fluid (DF). Concerning this DF, in the literature, one usually separates it into two energy components, i.e. the named dark matter and dark energy for recent reviews on dark energy please see \cite{ref:DEReview1,ref:DEReview2,ref:DEReview3,ref:DEReview4,ref:DEReview5,ref:DEReview6,ref:DEReview7}, or treats it as a unified dark fluid (UDF) \cite{ref:darkdeneracy,ref:XUCASS,ref:XUGCG, ref:XUMCG,ref:MCG,ref:GCG,ref:CG,ref:MCGall,ref:Lu,ref:LuMCG}. 

The most popular UDF models include the so-called Chaplygin gas (Cg) model, its extensions the gCg model and the modified Chaplygin gas (mCg) model and the model with a constant adiabatic sound speed (CASS), which have been discussed extensively in the literature \cite{ref:darkdeneracy,ref:XUCASS,ref:XUGCG, ref:XUMCG,ref:MCG,ref:GCG,ref:CG,ref:MCGall,ref:Lu,ref:LuMCG} where cosmic observations from SN Ia, BAO, CMB, Gamma-ray bursts and OHD data points have been used to test the viability of this UDF model and to constrain the model parameter space. Recently, tight constraints to the gCg, mCg and CASS models were obtained by using the data points from SN Ia, BAO and full CMB information from WMAP7 \cite{ref:XUGCG, ref:XUMCG,ref:XUCASS}. It was also pointed out that they are competitive cosmological models to $\Lambda$CDM model \cite{ref:XUGCG, ref:XUMCG,ref:XUCASS}.

For a perfect fluid with EoS $w(a)=p/\rho$, one can easily obtain the adiabatic sound speed $c^2_s(a)=\delta p/\delta \rho=\dot{p}/\dot{\rho}$ in the form of
\begin{equation}
c^2_s(a)=w(a)-\frac{1}{3}\frac{d \ln (1+w(a))}{d\ln a},
\end{equation}
which characterizes the propagation of linear perturbation of UDF. When a EoS is specified, one can derive the corresponding adiabatic sound speed, and vice versa. In fact, for a constant adiabatic sound speed case, the model has been investigated in Ref. \cite{ref:XUCASS}. A general case was also discussed in \cite{ref:Caplar}, where a relation $c^2_s=\alpha(-w)^\gamma$ was proposed. In this model, when $\gamma=1$, it reduces to the gCg model. As known for a gCg model, the EoS is given as
\begin{equation}
w(a)=-B_s/(B_s+(1-B_s)a^{-3(1+\alpha)}).
\end{equation}
It would be interesting to consider the EoS of gCg from another points of view. Let us consider the $\Lambda$CDM model at first, one can composite the cold dark matter and the vacuum energy into a UDF, then the coined UDF has EoS
\begin{equation}
w(a)=-\frac{\Omega_{\Lambda}}{\Omega_{\Lambda}+(1-\Omega_{\Lambda})a^{-3}}\label{eq:ULCDM}.
\end{equation}
Along this line, for a quintessence dark energy model with a constant EoS $w$, one can also coin a UDF with EoS
\begin{equation}
w(a)=-\frac{\Omega_{de0}}{\Omega_{de0}+(1-\Omega_{de0})a^{3w}}\label{eq:UQCDM}.
\end{equation}
Put it clearly, one can rewrite the Eqs. (\ref{eq:ULCDM}, \ref{eq:UQCDM}) into a uniformed form
\begin{equation}
w(a)=-\frac{1}{\beta a^{-n}+1},
\end{equation}
where $\beta=(1-B_s)/B_s$ or $\beta=(1-\Omega_{de0})/\Omega_{de0}$ ($\Omega_{de0}=\Omega_{\Lambda}$) and $n=3(1+\alpha)$ or $n=-3w$ ($w=-1$) for the gCg and quintessence ($\Lambda$CDM) model respectively. Then one can see that the gCg model is nothing but from a coined EoS from the $\Lambda$CDM model. So one would not be surprised that the currently available geometric observations from SN Ia, BAO, CMB cannot distinguish the $\Lambda$CDM model from the gCg model. This situation becomes more serious when one uses the derived values of the CMB shift parameters ($z_\ast$, $\l_a$ and $R$) based on $\Lambda$CDM model. This circular problem is the obvious drawback when the derived CMB shift parameters are used to test viability or constrain the model parameter space. 

Along this line, for generalization, we propose a {\it new} UDF model with the EoS given in the form of
\begin{equation}
w(a)=-\alpha/(\beta a^{-n}+1)\label{eq:neweos},
\end{equation}
where $\alpha$, $\beta$ and $n$ are three positive numbers. This EoS includes the gCg model as its special case when $\alpha=1$, $\beta=(1-B_s)/B_s$ and $n=3(1+\alpha)$. A novel character of this EoS is the fact that it gives more possibilities to deviate from the $\Lambda$CDM model in a simple way when $\alpha$ and $n$ deviate from $1$ and $3$ respectively. 

In this paper, we do not want to decompose the UDF energy component into the so-called dark matter and dark energy components. As we have pointed out in our previous papers \cite{ref:XUGCG, ref:XUMCG,ref:XUCASS} that the decomposition is not unique. Then unphysical effect would be introduced by decomposition. 

Then one can use the currently available cosmic observations to test the viability of this model and investigate the possible deviation from $\Lambda$CDM model. When one is doing the test, the more important is to avoid the circular problem. To satisfy this condition, we will use the SN Ia, BAO and full information of CMB from WMAP7 instead of the derived CMB shift parameters to test the viability of the new UDF model. 

This paper is structured as follows. In section \ref{sec:UDF}, the basic equations for UDF are shown which include the background and perturbation equations. In section \ref{sec:method}, the constraint method and results are presented. A summary is given in Section \ref{ref:conclusion}.

\section{Basic equations of Cosmology}   \label{sec:UDF}  

For a space with uniform curvature, the line element is given by the Friedmann-Robertson-Walker (FRW) metric
\begin{equation}
ds^{2}=-dt^{2}+a^{2}(t)\left[\frac{1}{1-kr^{2}}dr^{2}+r^{2}(d\theta^{2}+\sin^{2}\theta d\phi^{2})\right],
\end{equation}
where $k=0,\pm 1$ is the three-dimensional curvature and $a(t)$ is the scale factor. With our proposed new EoS (\ref{eq:neweos}), considering the energy conservation, one has the energy density $\rho_{U}$ for the UDF
\begin{eqnarray}
\rho_{U}&=&\rho_{U0}\exp\left\{-3\int^a_1\left[1+w(a)\right]d\ln a\right\}\nonumber\\
&=&\rho_{U0}\left(\frac{a^n+\beta}{1+\beta}\right)^{\frac{3\alpha}{n}}/a^3. \label{eq:rhoUDF}
\end{eqnarray}
Then one has the Friedmann equation from the Einstein equations
\begin{eqnarray}
H^{2}&=&H^{2}_{0}\left\{\Omega_{b}a^{-3}+\Omega_{r}a^{-4}+\Omega_{k}a^{-2}\right.\nonumber\\
&+&\left.\Omega_{U}\left[(a^n+\beta)/(1+\beta)\right]^{3\alpha/n}a^{-3}\right\}
\end{eqnarray}
where $H$ is the Hubble parameter with its current value $H_{0}=100h\text{km s}^{-1}\text{Mpc}^{-1}$, and $\Omega_{i}$ ($i=b,r,k$) are dimensionless energy parameters of baryon, radiation and effective curvature density respectively. Here $\Omega_{U}=1-\Omega_{b}-\Omega_{r}-\Omega_{k}$ is the dimensionless energy density for the UDF. In this paper, we only consider the spatially flat FRW universe.

The perturbation equations for UDF should be investigated when one considers the effects on the CMB anisotropic power spectrum. Here, we will not consider any direct interaction between UDF with any other energy components. It means that the UDF interacts with the rest of matter purely through gravity. By the assumption of pure adiabatic contribution to the perturbations, the speed of sound for UDF is written in the form 
\begin{eqnarray}
c^{2}_{s}(a)&=&w(a)-\frac{1}{3}\frac{d \ln (1+w(a))}{d\ln a}\nonumber\\
&=&-\alpha+\frac{\beta n}{(1-\alpha)a^n+\beta}-\frac{\beta(n-3\alpha)}{a^n+\beta},
\label{eq:cs2}
\end{eqnarray}
which should be in the range $[0,1]$ to keep the perturbation evolution stable and to avoid casual problem. However, it is not easy to give the parameter space analytically where $c^2_s$ is a small positive number for any value of scale factor $a$. We will use the currently available cosmic observations to constrain the possible parameter space. We will explain the technical issue in the next section. 

In the synchronous gauge, using the conservation of energy-momentum tensor $T^{\mu}_{\nu;\mu}=0$, one has the perturbation equations of density contrast and velocity divergence for UDF
\begin{eqnarray}
\dot{\delta}_{U}&=&-(1+w)(\theta_{U}+\frac{\dot{h}}{2})-3\mathcal{H}(c^{2}_{s}-w)\delta_{U}\\
\dot{\theta}_{U}&=&-\mathcal{H}(1-3c^{2}_{s})\theta_{U}+\frac{c^{2}_{s}}{1+w}k^{2}\delta_{U}-k^{2}\sigma_{U}
\end{eqnarray}
following the notation of Ma and Bertschinger \cite{ref:MB}. The shear perturbation $\sigma_{U}=0$ is assumed and the adiabatic initial conditions are adopted in our calculation.

\section{Constraint method and results}\label{sec:method}

\subsection{Method and data points}

We test the viability of this model by performing the observational constraints on parameter space via the Markov Chain Monte Carlo (MCMC) method which is contained in a publicly available cosmoMC package \cite{ref:MCMC}, including the CAMB \cite{ref:CAMB} code to calculate the theoretical CMB power spectrum. We modified the code for the UDF model with its perturbations included. The following $8$-dimensional parameter space  is adopted
\begin{equation}
P\equiv\{\omega_{b},\Theta_{S},\tau, \alpha,\beta,n,n_{s},\log[10^{10}A_{s}]\}
\end{equation}
where $\omega_{b}=\Omega_{b}h^{2}$ is the physical baryon density, $\Theta_{S}$ (multiplied by $100$) is the ratio of the sound horizon and angular diameter distance, $\tau$ is the optical depth, $\alpha$, $\beta$ and $n$ are three newly added model parameters related to UDF, $n_{s}$ is scalar spectral index, $A_{s}$ is the amplitude of of the initial power spectrum. Please notice that the current dimensionless energy density of UDF $\Omega_{U}$ is a derived parameter in a spatially flat ($k=0$) FRW universe. So, it is not included in the model parameter space $P$. The pivot scale of the initial scalar power spectrum $k_{s0}=0.05\text{Mpc}^{-1}$ is used. We take the following priors to model parameters: $\omega_{b}\in[0.005,0.1]$, $\Theta_{S}\in[0.5,10]$, $\tau\in[0.01,0.8]$, $\alpha\in[0,5]$, $\beta\in[0, 1]$, $n\in[0,5]$, $n_{s}\in[0.5,1.5]$ and $\log[10^{10}A_{s}]\in[2.7, 4]$. In addition, the hard coded prior on the comic age $10\text{Gyr}<t_{0}<\text{20Gyr}$ is imposed. Also, the weak Gaussian prior on the physical baryon density $\omega_{b}=0.022\pm0.002$ \cite{ref:bbn} from big bang nucleosynthesis and new Hubble constant $H_{0}=74.2\pm3.6\text{kms}^{-1}\text{Mpc}^{-1}$ \cite{ref:hubble} are adopted. 

As is shown in Eq. (\ref{eq:cs2}), the expression of $c^2_s$, which contains model parameters $\alpha$, $\beta$, $n$ and scale factor $a$, is complicated. Giving an explicit range of model parameters to keep $c^2_s$ nonnegative is really difficult. Maybe, in some senses, it is impossible. To circumvent the problem, we take the code as a black box, and hard code the condition $1\ge c^2_s(a)\ge0$ in the sampling process. It means that, in very sampling, the values of $\alpha$, $\beta$ and $n$ can go to the next calculation stage only if they give the values of $c^2_s$ satisfying the constrained condition for every value of the scale factor $a$. We can check the evolution of $c^2_s$ with respect to scale factor $a$ once the final result is obtained. As is shown in the following section, please see Figure \ref{fig:wdcs}, the strategy really works.

The distribution of parameter space is given by calculating the total likelihood $\mathcal{L} \propto e^{-\chi^{2}/2}$, here the $\chi^{2}$ is given as
\begin{equation}
\chi^{2}=\chi^{2}_{CMB}+\chi^{2}_{BAO}+\chi^{2}_{SN}.
\end{equation}
The CMB data include temperature and polarization power spectrum from WMAP $7$-year data \cite{ref:lambda}. The additional geometric constraint comes from standard ruler BAO and standard candle SN Ia. For BAO, the values $\{r_{s}(z_{d})/D_{V}(0.2),r_{s}(z_{d})/D_{V}(0.5)\}$ and their inverse covariant matrix \cite{ref:BAO} are used. To use the BAO information, one needs to know the sound horizon at the redshift of drag epoch $z_{d}$. Usually, $z_{d}$ is obtained by using the accurate fitting formula \cite{ref:EH} which is valid if the matter scalings $\rho_{b}\propto a^{-3}$ and $\rho_{c}\propto a^{-3}$ are respected. Obviously, it is not true in our case. So, we find $z_{d}$ numerically from the following integration
\cite{ref:Hamann}
\begin{eqnarray}
\tau(\eta_d)&\equiv& \int_{\eta}^{\eta_0}d\eta'\dot{\tau}_d\nonumber\\
&=&\int_0^{z_d}dz\frac{d\eta}{da}\frac{x_e(z)\sigma_T}{R}=1
\end{eqnarray}   
where $R=3\rho_{b}/4\rho_{\gamma}$, $\sigma_T$ is the Thomson cross-section and $x_e(z)$ is the fraction of free electrons. Then the sound horizon is
\begin{equation}
r_{s}(z_{d})=\int_{0}^{\eta(z_{d})}d\eta c_{s}(1+z).
\end{equation}   
where $c_s=1/\sqrt{3(1+R)}$ is the sound speed. We use the substitution \cite{ref:Hamann}
\begin{equation}
d_z\rightarrow d_z\frac{\hat{r}_s(\tilde{z}_d)}{\hat{r}_s(z_d)}r_s(z_d),
\end{equation}
to obtain unbiased parameter and error estimates, where $d_z=r_s(\tilde{z}_d)/D_V(z)$, $\hat{r}_s$ is evaluated for the fiducial cosmology of Ref. \cite{ref:BAO}, and $\tilde{z}_d$ is obtained by using the fitting formula \cite{ref:EH} for the fiducial cosmology. Here $D_V(z)=[(1+z)^2D^2_Acz/H(z)]^{1/3}$ is the 'volume distance' with the angular diameter distance $D_A$. 
The $580$ Union2.1 data with systematic errors are also included \cite{ref:Union21}. For the detailed description of SN, please see Refs. \cite{ref:Union21,ref:Xu}.

\subsection{Fitting Results and Discussion}

We generate $8$ independent chains in parallel and stop sampling by checking the worst e-values [the
variance(mean)/mean(variance) of 1/2 chains] $R-1$ of the order $0.01$. The fitting results of the model parameters and derived parameters are shown in Table. \ref{tab:results}, where the mean values with $1\sigma$, $2\sigma$ and $3\sigma$ regions from the combination WMAP+BAO+SN are listed. Correspondingly, the contour plots are shown in Figure \ref{fig:contour}. For comparison, using the same data combination, we show the corresponding results for $\Lambda$CDM model also in Table \ref{tab:results}. For intimately touching to the $\Lambda$CDM model, the $\alpha=1$ case is also explored for the same data combination.

\begin{widetext}
\begin{center}
\begin{table}
\begin{tabular}{cccc}
\hline\hline Prameters & NUDF Mean with errors & NUDF($\alpha=1$) Mean with errors & $\Lambda$CDM Mean with errors \\ \hline
$\Omega_b h^2$ & $    0.0230_{-    0.000567-    0.00111-    0.00170}^{+    0.000562+    0.00112+    0.00176}$ & $    0.0228_{-    0.000561-    0.001081-    0.00160}^{+    0.000554+    0.00111+    0.00173}$ & $    0.0226_{-    0.000509-    0.00101-    0.00145}^{+    0.000514+    0.00100+    0.00150}$\\
$\Omega_{DM} h^2$ & - & - & $    0.110_{-    0.00368-    0.00711-    0.0110}^{+    0.00367+    0.00736+    0.0113}$\\
$\theta$ & $    1.0485_{-    0.00253-    0.00491-    0.00763}^{+    0.00248+    0.00502+    0.00747}$ & $    1.0485_{-    0.00257-    0.00503-    0.00784}^{+    0.00251+    0.00484+    0.00736}$ & $    1.0398_{-    0.00252-    0.00506-    0.00755}^{+    0.00253+    0.00493+    0.00755}$ \\
$\tau$ & $    0.0903_{-    0.00721-    0.0239-    0.0388}^{+    0.00657+    0.0250+    0.0436}$ & $    0.0905_{-    0.00730-    0.0234-    0.0380}^{+    0.00639+    0.0249+    0.0423}$ & $    0.0927_{-    0.00719-    0.0235-    0.0379}^{+    0.00647+    0.0246+    0.0452}$ \\
$\alpha$ & $    1.000546_{-    0.000825-    0.00143-    0.00206}^{+    0.000863+    0.00229+    0.00427}$ & - & -\\
$\beta$ & $    0.277_{-    0.0256-    0.0475-    0.0679}^{+    0.0257+    0.0557+    0.0890}$ & $    0.280_{-    0.0261-    0.0476-    0.0681}^{+    0.0258+    0.0544+    0.0887}$ & -\\
$n$ & $    3.00487_{-    0.00487-    0.00487-    0.00487}^{+    0.00108+    0.00787+    0.0158}$ & $    3.00410_{-    0.00410-    0.00410-    0.00410}^{+    0.000867+    0.00649+    0.0135}$ & -\\
$n_s$ & $    0.988_{-    0.0150-    0.0290-    0.0453}^{+    0.0151+    0.0308+    0.0502}$ & $    0.982_{-    0.0149-    0.0285-    0.0400}^{+    0.0146+    0.0304+    0.0488}$ & $    0.971_{-    0.0117-    0.0232-    0.0342}^{+    0.0119+    0.0236+    0.0370}$\\
$\log[10^{10} A_s]$ &$    3.0823_{-    0.0346-    0.0671-    0.0998}^{+    0.0346+    0.0708+    0.110}$ & $    3.0816_{-    0.0346-    0.0652-    0.0949}^{+    0.0342+    0.0699+    0.105}$ & $    3.0853_{-    0.0349-    0.0664-    0.0978}^{+    0.0344+    0.0703+    0.106}$\\
$\Omega_U(\Omega_\Lambda)$ & $    0.956_{-    0.00165-    0.00338-    0.00530}^{+    0.00164+    0.00323+    0.00488}$& $    0.956_{-    0.00168-    0.00331-    0.00512}^{+    0.00168+    0.00320+    0.00477}$ & $    0.742_{-    0.0160-    0.0333-    0.0512}^{+    0.0162+    0.0296+    0.0440}$\\
$Age/Gyr$ & $   13.664_{-    0.112-    0.220-    0.333}^{+    0.112+    0.221+    0.330}$ & $   13.684_{-    0.111-    0.217-    0.334}^{+    0.112+    0.218+    0.335}$ & $   13.718_{-    0.105-    0.201-    0.312}^{+    0.106+    0.208+    0.311}$\\
$\Omega_b$($\Omega_m$) & $    0.044_{-    0.0016-    0.0032-    0.0049}^{+    0.0017+    0.0034+    0.0053}$ & $    0.044_{-    0.0017-    0.0032-    0.0048}^{+    0.0017+    0.0033+    0.0051}$ & $    0.258_{-    0.0162-    0.0296-    0.0439}^{+    0.0160+    0.0333+    0.0512}$ \\
$z_{re}$ & $   10.544_{-    1.167-    2.403-    3.789}^{+    1.191+    2.329+    3.670}$  & $   10.616_{-    1.180-    2.299-    3.605}^{+    1.192+    2.340+    3.516}$ & $   10.891_{-    1.192-    2.371-    3.578}^{+    1.189+    2.338+    3.603}$ \\
$H_0$ & $   72.493_{-    1.579-    3.0947-    4.596}^{+    1.587+    3.140+    4.900}$ & $   72.211_{-    1.546-    2.983-    4.461}^{+    1.556+    3.118+    4.898}$& $   71.629_{-    1.435-    2.845-    4.0537}^{+    1.446+    2.783+    4.381}$\\
\hline\hline
\end{tabular}
\caption{The mean values of model parameters with $1-3\sigma$ errors from the combination WMAP+BAO+SN.}\label{tab:results}
\end{table}
\end{center}
\end{widetext}

\begin{widetext}
\begin{center}
\begin{figure}[tbh]
\includegraphics[width=15cm]{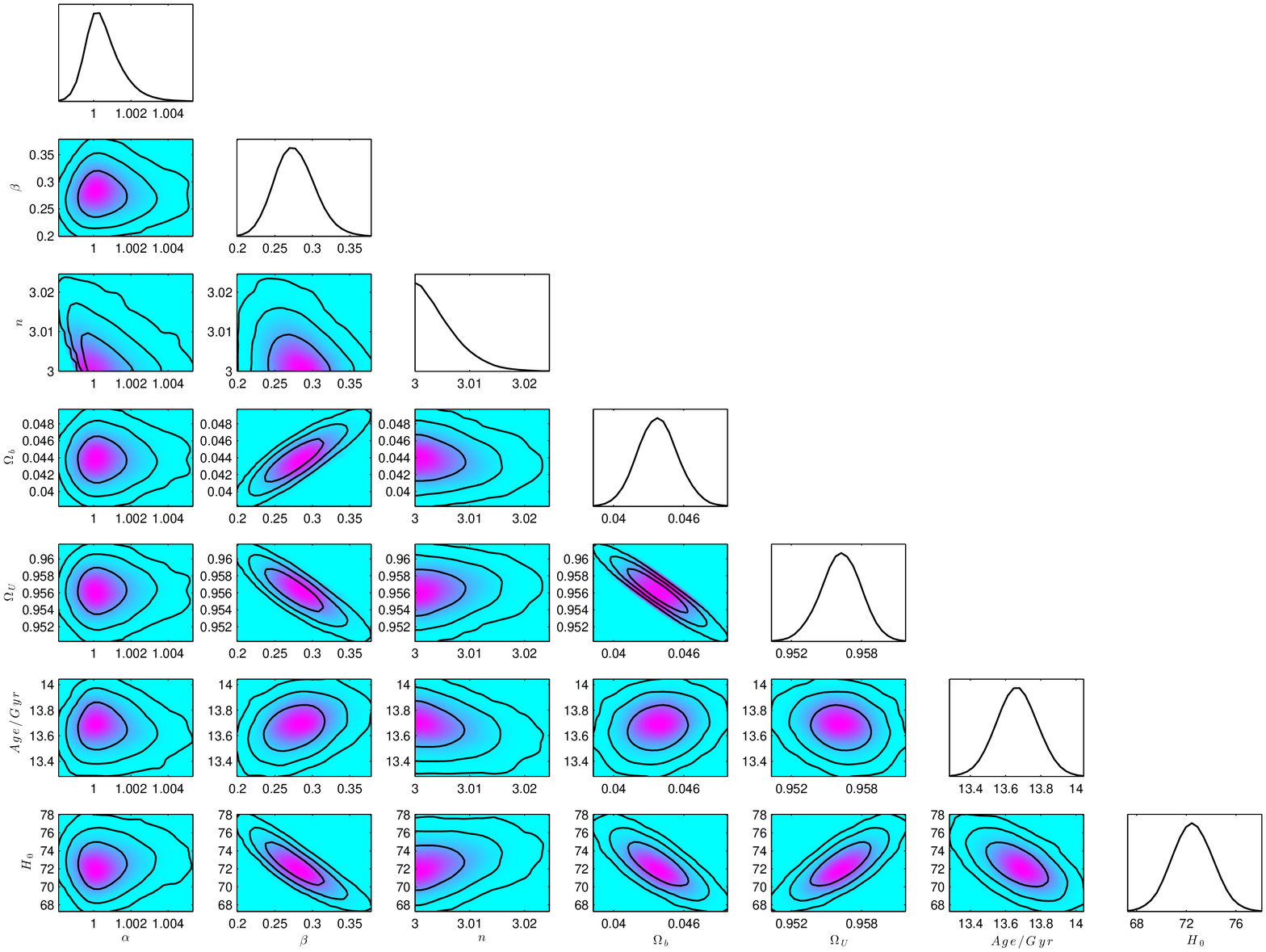}
\includegraphics[width=15cm]{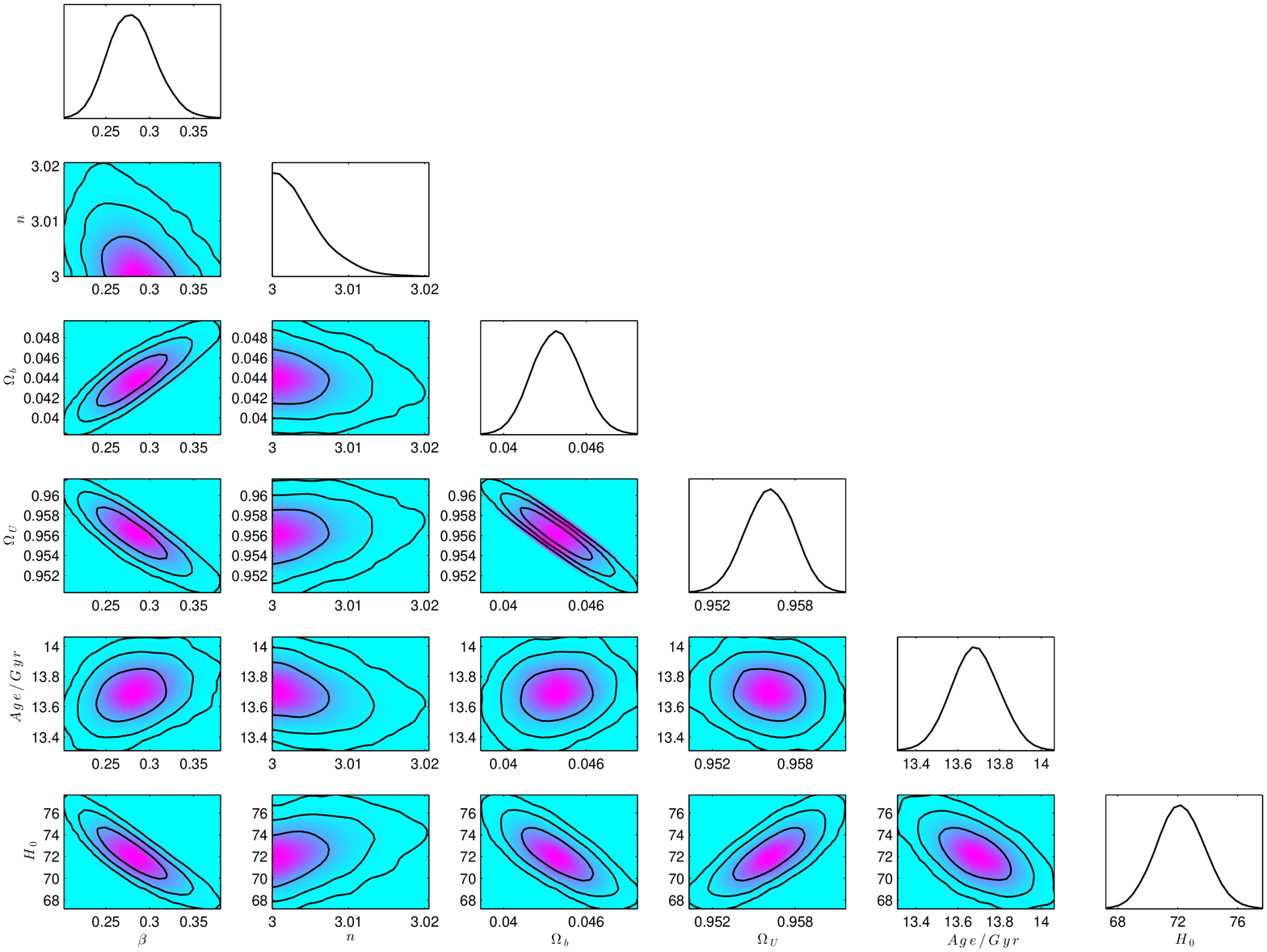}
\caption{The 2-dimensional contours and 1-dimensional probability distribution of model parameters with $1-3\sigma$ errors. Bottom panel  show the case where the mode parameter $\alpha$ is fixed to $1$.}\label{fig:contour}
\end{figure}
\end{center}
\end{widetext}

From the Table \ref{tab:results} and Figure \ref{fig:contour}, one can clearly see that a tight constraint to the model parameters $\alpha$, $\beta$ and $n$ is obtained. The results indicate that the proposed UDF model do not deviate from $\Lambda$CDM model at the background level distinctly. The values of $\alpha$ and $\beta$ are very close to the $\Lambda$CDM model limit $1$ and $3$ respectively.  

Using the mean values of model parameters, we plot the evolution of EoS $w(a)$ and the speed of sound $c^2_s(a)$ for UDF with respect to the scale factor $a$ in Figure \ref{fig:wdcs}. From the left panel of Figure \ref{fig:wdcs}, the UDF behaves like cold dark matte at early epoch and dark energy at late epoch. The right panel of Figure \ref{fig:wdcs} shows that the value of the sound speed $c^2_s(a)$ of UDF is small positive number and varies with scale factor. The small values of $c^2_s(a)$ make it possible to form large scale structures in our universe. Also, the positivity of $c^2_s(a)$ is really guaranteed by the 'filter' in sampling. One can read the fact that the $c^2_s(a)$ for UDF evolves with respect to the scale factor $a$ but  that is constant and zero for the $\Lambda$CDM model from the Figure \ref{fig:wdcs}. It implies that the observational effects that can be used to discriminate the UDF model from $\Lambda$CDM model are due to the sound speed and EoS. The large structure formation information and late time integrated Sachs-Wolfe (ISW) effect can be combined with background evolution data sets to discriminate the models. However, to use the observed large structure information, one has to investigate the large scale structure formation process in UDF model which is out the range of the current paper.  

\begin{widetext}
\begin{center}
\begin{figure}[tbh]
\includegraphics[width=8.3cm]{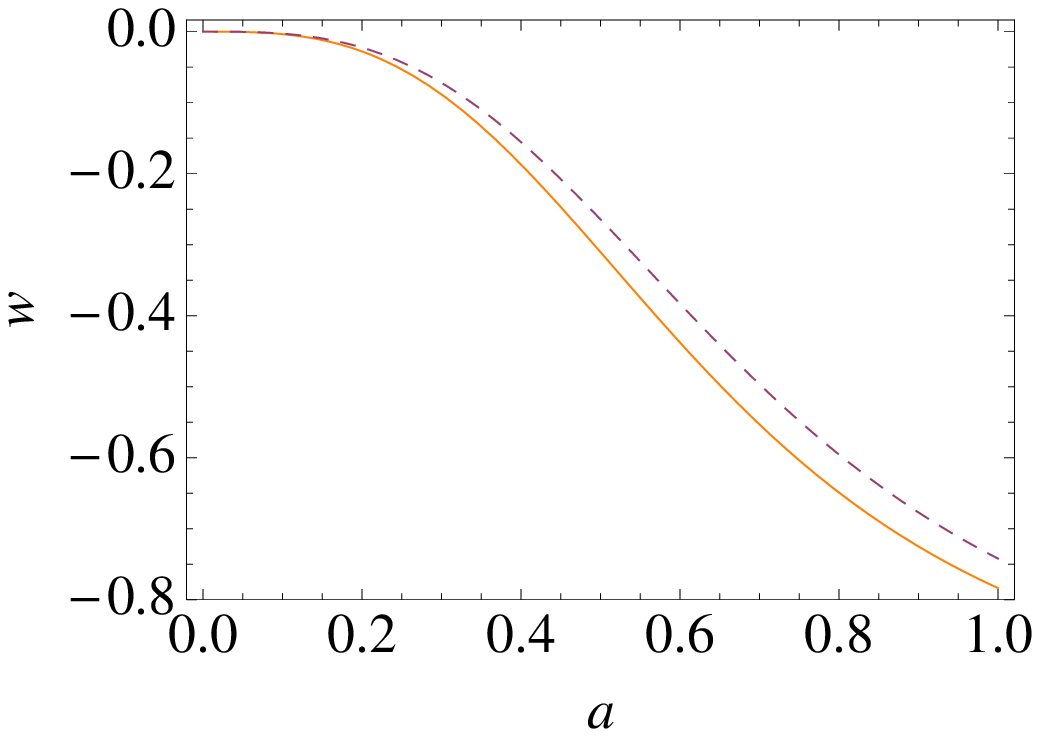}
\includegraphics[width=8.8cm]{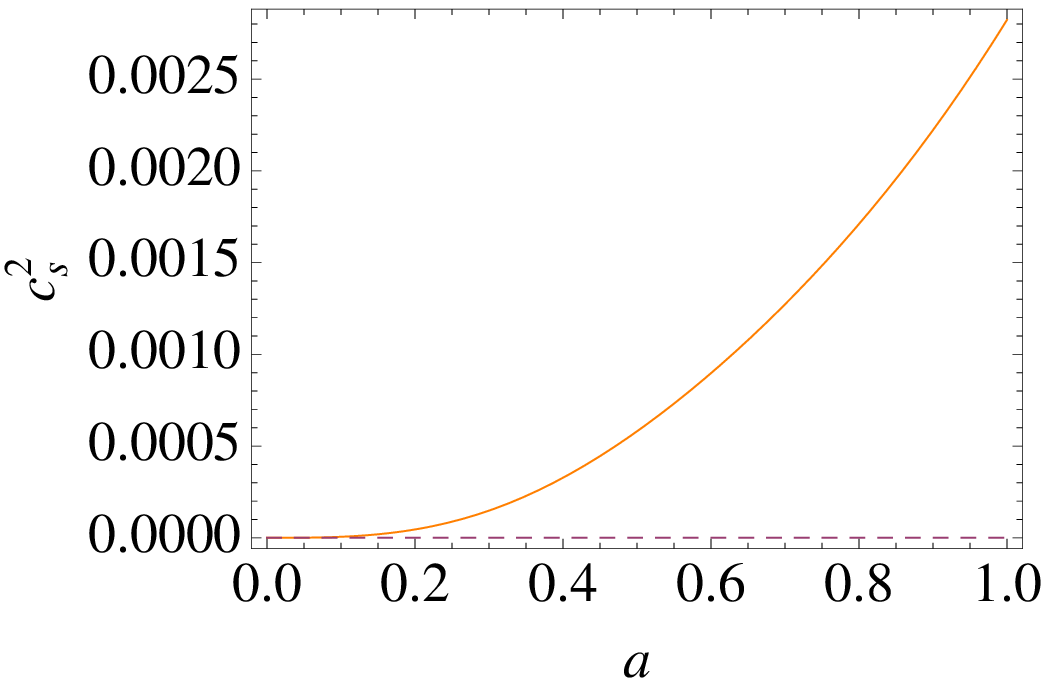}
\caption{The evolutions of EoS $w(a)$ and sound speed $c^2_s(a)$ for UDF (orange solid lines) and $\Lambda$CDM (pink dashed lines) with respect to the scale factor $a$, where the mean values of the model parameters are adopted.}\label{fig:wdcs}
\end{figure}
\end{center}
\end{widetext}

For comparison, we show the CMB $C^{TT}_{l}$ power spectrum v.s. the multiple momentum $l$ in Figure \ref{fig:cls} where the mean values of the relevant cosmological and model parameters are adopted from the Table \ref{tab:results} . It is clearly that the curves of the power spectra of UDF match up to that of $\Lambda$CDM model. The tiny difference comes from the hight of the fist peak and the trough at the large scale ($l<100$). The difference can be understood easily, if one writes down the correspondence $\beta=(1-\Omega_{de0})/\Omega_{de0}$. Large values of $\beta$ mean large ratio of the matter and dark energy. Large values of $\Omega_{m0}=1-\Omega_{de0}$ make the equality time of matter and radiation earlier if they evolve with the same scaling law, say $\propto a^{-3}$, then the first peak of the CMB power spectrum is shifted to a higher position. At the large scale ($l<100$), the difference is due to the tiny deviation from the scaling law $\propto a^{-3}$, i.e. through the integrated Sachs-Wolfe (ISW) effect.   

\begin{widetext}
\begin{center}
\begin{figure}[tbh]
\includegraphics[width=16cm]{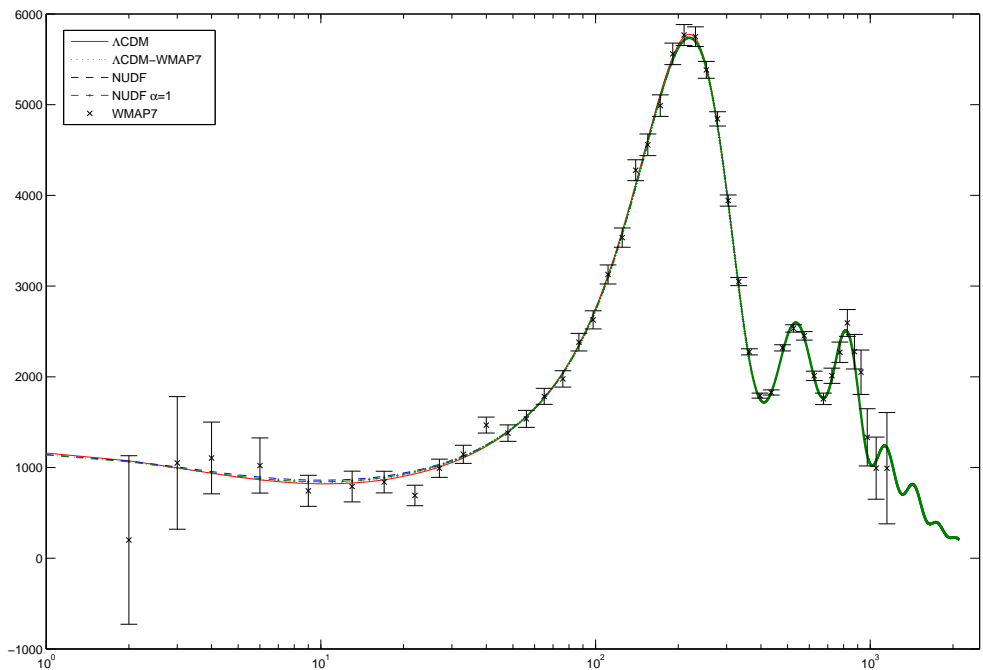}
\caption{The CMB $C^{TT}_l$ power spectrum v.s. multiple moment $l$. The relevant model parameters are adopted to the mean values as listed in Table \ref{tab:results} for NUDF and $\Lambda$CDM model, where the black dots with error
bars denote the observed data with their corresponding uncertainties from WMAP $7$-year results, the blue dashed line is for NUDF with $\alpha$ free, and the green dashed line for NUDF with $\alpha=1$, the red solid line is for $\Lambda$CDM model with mean values for the same data points combination. And the red doted line is for $\Lambda$CDM model with mean values taken from \cite{ref:wmap7} with WMAP+BAO+$H_0$ constraint results.}\label{fig:cls}
\end{figure}
\end{center}
\end{widetext}

\section{Summary} \label{ref:conclusion} 

In summary, we proposed a new unified dark fluid model $w(a)=-\alpha/(\beta a^{-n}+1)$ which is inspired by a combination of the cold dark matter and the cosmological model constant. In this simple form, one can see that the EoS of the gCg model is nothing but a coined one from that of $\Lambda$CDM model. And the gCg model is a special case of our proposed one. The new unified model gives more possibilities to deviated from the $\Lambda$CDM model in the unified dark fluid case when the model parameters $\alpha$ and $n$ take different values $a$ and $3$ respectively. To test the viability of this new UDF model and detect the possible deviation from $\Lambda$CDM model, we perform a global fitting by using the MCMC method with a combination of the full CMB, BAO and  SN Ia data points. The results show that the new UDF model fits the observational data sets as well as that of $\Lambda$CDM model. And, one does not see any deviation from $\Lambda$CDM model in $3\sigma$ confidence level under the currently available data sets. But I would like to point out that the new proposed UDF model can give a non-zero sound speed. As a contrast, the coined EoS from $\Lambda$CDM model has zero sound speed. Base on this point, we expect large scale structure formation information can distinct the new UDF model from $\Lambda$CDM model.  

\section{Acknowledgements} L. Xu's work is supported by the Fundamental Research Funds for
the Central Universities (DUT11LK39) and in part by NSFC under the Grants No. 11275035.

\end{document}